\documentstyle[11pt]{article}

\renewcommand{\phi}{\varphi}
\newcommand{\IT}{{\rm IT}}

\newcommand{\term}{{\rm TERM}}
\newcommand{\RHU}{{\rm RHU}}
\newcommand{\RHB}{{\rm RHB}}

\newcommand{\calL}{{\cal L}}

\renewcommand{\>}{\rangle}
\renewcommand{\aa}{\underline}
\newcommand{\oo}{\overline}

\newcommand{\LIM}{{\rm LIM}}

\newcommand{\dfrac}{\ddd\frac}

\newcommand{\rt}{\rightarrow}
\newcommand{\dlim}{\ddd\lim}
\newcommand{\ddd}{\displaystyle}

\begin{document}

\begin{center}
{\Large {\bf A Logical Framework for Convergent Infinite Computations}} 
\footnote[1]{{\small The research is supported by the National 973 Project
of China under the grant number G1999032701 and by the National Science
Foundation of China.}}\\[0.5cm]
{Wei Li$^\dag$\footnote[2]{{\small Email: \{liwei, slma,
kexu\}@nlsde.buaa.edu.cn.}}, Shilong Ma$^{\dag,2}$, Yuefei Sui$^\ddag$%
\footnote[3]{{\small Email: suiyyff@sina.com.}}, and Ke Xu$^{\dag,2}$}\\[0pt]
{\ {}$^\dag$ Department of Computer Science, Beijing University of
Aeronautics and Astronautics, Beijing 100083, CHINA}\\[0pt]
{\ {}$^\ddag$ Institute of Computing Technology, Chinese Academy of Sciences}%
\\[0pt]
{Beijing 100080, CHINA}
\end{center}

\bigskip

\begin{minipage}{12.5cm} 
{\small\footnotesize {\bf  Abstract}: \ Classical computations can not capture the essence of infinite computations 
very well. 
This paper will focus on a class of infinite computations called {\sl convergent infinite computations}. 
A logic for convergent infinite computations is 
proposed by extending first order theories using Cauchy sequences, which has 
stronger expressive power than the first order logic. A class of fixed 
points characterizing the logical properties of the limits can be 
represented by means of infinite-length terms defined by Cauchy sequences. We will show that the 
limit of sequence of first order theories can be defined in terms of 
distance, similar to the $\epsilon-N$ style definition of limits in real analysis. On the basis of infinitary 
terms, a computation model for convergent infinite computations is proposed.
Finally, the interpretations of logic programs are extended by introducing real 
Herbrand models of logic programs and a sufficient condition for computing 
a real Herbrand model of Horn logic programs using convergent infinite computation is given.

\noindent{\bf Keywords: } Infinite Computation, Distance, Limit, Convergent infinite computation.}
\end{minipage}

\bigskip

\noindent{\bf 1.Introduction}

\smallskip

The need of studying infinite computations has been emphasized in recent
years, e.g., see (Vardi and Wolper, 1994). By infinite computations, one
means the computations done by some programs that create non-terminating
processes or very long time running processes. For such programs, the
computations done by them usually go through infinite sequences of running
states (or configurations), unlike finite computations in which only finite
sequences of running states are involved.

Furthermore, over computer networks there is a very large family of
computations which are carried out very long time (approximately treated as
infinite time) and need constantly to interact with other processes and
access some huge sets of external data over networks (approximately treated
as infinite sets of data). For example, various procedures for knowledge
discovery from databases over Internet do such computations.

\bigskip \noindent{\it 1.1.\ Convergent infinite computations}

\smallskip In the above-mentioned family, there is a large class of infinite
computations that have the following characteristics: (1) They constantly
access some huge sets of external data during the run time, and (2) the
infinite sequences of running states, which they go through, are convergent
to some certain limits as the time goes to the infinity. Such computations
will be called {\sl convergent infinite computations} in this paper.

In the following, we will focus on convergent infinite computations, and
establish a logical framework for them. This requires us to study the computational
behaviors from the point of view of analyzing long time changes, because
such computations depend fundamentally on the nature of the long time
changes and classical computations may not capture their essence well.

As well known, a computation can be expressed by a first order theory. We
will give a framework for convergent infinite computations by expressing
infinite computations with sequences of first order theories. Our approach
is based on the study of the sequences of first order theories and their
limits (Li, 1992). The problems of infinite computations are reduced to
that of sequences of first order theories and their limits. The concept of
limit of a sequence of first order theories in (Li, 1992), different from
the previous concepts of the limit involved in computer science and
mathematical logic, is used to characterize that some theory is infinitely
approached but maybe never is reached. 

We will discuss a class of ideal long time changes, i.e., long time changes
with some \lq\lq continuous" nature, by extending first order theories and
domains using Cauchy sequences. The extension is similar to that of rational
numbers to real numbers. We will show that the semantic interpretations of
first order theories are enriched by extending domains using Cauchy
sequences. In addition, we show that the limit of sequence of first order
theories can be defined in terms of distance, similar to the $\epsilon-N$
style definition of limits in real analysis.

In sections 2,3 and 4, based on the study on sequences of first order
theories and their limits, a logic for convergent infinite computations is
proposed by extending first order theories using Cauchy sequences, which has
stronger expressive power than the first order logic. A class of fixed
points characterizing the logical properties of the limits can be
represented by means of infinite length terms defined by Cauchy sequences.
Furthermore, we study the relations between the convergence of theory
sequences and the convergence of model sequences to characterize the limits of formal theory
sequences from both proof-theoretical and model-theoretical approaches.
We give a formal computation model on the infinitary terms in section 5.
Finally, the interpretations of logic programs are extended by introducing real 
Herbrand models of logic programs and a sufficient condition for computing 
a real Herbrand model of Horn logic programs using convergent infinite computation is given.

\bigskip {\it 1.2.\ Comparison with related work} \smallskip 

\begin{enumerate}
\item In the infinitary logic (Vardi and Wolper, 1994) and (Abiteboul, Vardi
and Vianu, 1995) there are formulas of infinite length, but there is no term
of infinite length. We give a logic for the terms of infinite length and
discuss the convergence problem of logical theory sequences. The logic for
the terms of infinite length is different from the infinite logic in
expressiveness and logic properties. For example, {\it being finite} can be
expressed in the infinite logic, and {\it a fix point} of a monotonic
function in a complete lattice can be expressed in the logic for the
infinite length terms. The compactness theorem does not hold in the infinite
logic, but does in the logic for the terms of infinite length.

\item In real machines (Blum, Shub and Smale, 1989, 1998) and analytic
machines (Chadzelek and Hotz, 1999), a computation model for real numbers is
established to characterize continuous computations. We discuss the
computation problem on the terms of infinite length (strings of infinite
length over an infinite alphabet) to characterize \lq\lq continuous"
symbolic computations. 

\item To study the approximation problem of inductive logic programming and
machine learning, the distance of Herbrand interpretations is discussed
(Nienhuys-Cheng, 1997, 1998). The approximation concept in this paper,
similar to the approximation concept in real analysis, is more general, and
is used to characterize that some theory is infinitely approached but may
never be reached. Usually, the approximation sequences are non-monotonic.
The semantic interpretations of first order theories are enriched by
extending Herbrand universe to real Herbrand universe using Cauchy sequences.
\end{enumerate}

{\bf Notation} Our notation is standard. The notation in deductive databases follows 
from Dahr (1997). We use $%
\Gamma$ to denote theories or logic programs, and $\rho$ to denote the
distances between terms, formulas, theories and logic programs. We use $%
\omega$ to denote the set of all the natural numbers, $i,j,k,m,n$ to denote
the natural numbers, $f,g,h$ to denote functions, $p,q$ to denote
predicates, and $t,r,s$ to denote terms, $x,y,z$ to denote variables, $%
\phi,\psi$ to denote the formulas, and $R$ to denote the relations as in
relational databases.

\bigskip

\noindent{\bf 2. The Cauchy sequences of terms.}

\smallskip We pay attention to not only the convergent infinite computations
which runs in a long run, but also the possible outputs of such
computations. To research such outputs of convergent infinite computations,
we should have a way to represent such outputs.

In description logic, a recursive definition such as $X=f(X)$ may have many
solutions, where $f$ is a monotonic operator. By Tarski's theorem, there is
a unique least fixed solution $A_0$ and a unique greatest fixed solution $%
A_1 $ to the definition. What kinds of description logical properties $A_0$
should have is a very interesting problem in description logic.

In mathematical analysis, given a continuous function $f$ and a real number $%
x,$ we cannot compute $f(x)$ directly. Instead, we can approximate $f(x)$ by
computing $f(x^{\prime})$ for some rational number $x^{\prime}$ close enough
to $x.\ f(x)$ can be taken as a result of infinite computations $%
(f(x_1),f(x_2),...,f(x_k),...),$ where the distance between $x$ and $x_k$ is
less than $\displaystyle\frac{1}{k}.$

\bigskip

In the following two sections we shall give a logic framework for the
convergent infinite computations based on the distance defined on the terms
of a language. In this section we shall give the distance on the terms of a
language which is basically equivalent to the distance defined on trees. By
the defined distance, we extend finite terms to infinitary terms. Such
extension is similar to the extension of rational numbers to real numbers,
but there are some differences between the two extensions. In the next
section we shall give a logic for the infinitary terms.

\bigskip

Before giving the definition of the infinite terms, we first define a
distance on the terms of some language.

Let ${\cal L}$ be a language consisting of constant symbols, variable
symbols, predicate symbols and function symbols; the logic connectives: $%
\lnot, \land, \exists.$ We shall use $c,d,...$ to denote the constant
symbols, $x,y,z,...$ to denote the variable symbols, $f,g,h,...$ the
function symbols and $p,q,...$ the predicate symbol.

A string $t$ of symbols in ${\cal L}$ is a {\it term} if

(i)\ $t$ is a constant symbol $c$, or

(ii)\ $t$ is a variable symbol $x$, or

(iii)\ $t$ is $f(t_1,...,t_n),$ where $f$ is an $n$-ary function and $%
t_1,...,t_n$ are terms.

\noindent We call the terms as the finitary terms, denote the set of all the
finitary terms by ${\rm FT}_\calL$ (we usually omit ${\cal L}$ when no
confusion occurs).

We define a distance on ${\rm FT}.$ Nienhuys-Cheng (Nienhuys-Cheng, 1997)
proposed a distance on terms and formulas. Here, we give a ramified
definition of a distance on ${\rm FT}.$

\vspace{2mm}

\noindent {\bf Definition 2.1.} Let $f$ and $g$ be an $n^{\prime}$-ary and
an $n$-ary function symbols, respectively. The distance $\rho: {\rm FT}
\times {\rm FT} \rightarrow \left\{\displaystyle\frac{1}{m}\mid m\in
\omega\right\}$ is defined as follows.

(i)\ $\rho(t, t)=0$, for any $t \in {\rm FT}$.

(ii)\ If $f \not= g$, then $\rho(f(t^{\prime}_1,...,t^{\prime}_{n^{%
\prime}}), g(t_1,..., t_n))=1$.

(iii)\ $\rho(f(t^{\prime}_1,..., t^{\prime}_{n^{\prime}}),
f(t_1,...,t_{n^{\prime}}))=\displaystyle\frac{\max\{\rho(t^{\prime}_i,
t_i)\mid 1\leq i\leq n^{\prime}\}}{\max\{\rho(t^{\prime}_i, t_i)\mid 1\leq
i\leq n^{\prime}\}+1}.$

{\bf Definition 2.2.} Let $t=\{t_k \mid k \in \omega\}$ be an infinite
sequence of terms. If for any number $m>0$, there exists an integer $K >0$
such that $\rho(t_k, t_j)< \displaystyle\frac{1}{m}$ for any $k,j \ge K,$
then $t$ is called a {\it Cauchy term sequence,}\ or simply, a {\it Cauchy
sequence,}\ an {\it infinitary term,}\ denoted by $t=\displaystyle%
\lim_{k\rightarrow \infty} t_k.$ Let ${\rm IT}_\calL,$ simply ${\rm IT},$ be
the set of all the Cauchy term sequence in ${\cal L}.$

{\bf Remark.} In what follows, we still use $t=\{t_k \mid k \in \omega\}$ to denote 
a Cauchy sequence $t=\displaystyle\lim_{k\rightarrow \infty} t_k$ when no confusion occurs.

\bigskip The definition of the distance on terms and formulas is a little
different from the one given by Nienhuys-Cheng in that the value which the
distance can take has a simple form $\displaystyle\frac{1}{m}$ for some
natural number. Such a distance is used to define the distance between two
trees in graph theory. Every term $t$ can be taken as a tree $T_t.$ For
example, $t=f(t_1,...,t_n),$ the tree $T_t$ has a root with symbol $f$ and $%
n $-many children $T_{t_1},...,T_{t_n}.$ We say that two terms $t$ and $%
t^{\prime}$ are the same to depth $m$ if $T_t$ and $T_{t^{\prime}}$ are the
same to depth $m$. The distance defined here have the basic properties that
the distance defined by Nienhuys-Cheng has. In the following, we give some
facts stated as propositions without proofs, since they can be proved easily.

{\bf Proposition 2.3.} (i)\ $t$ and $t^{\prime}$ are the same to depth $m$
if and only if $\rho(t,t^{\prime})\leq \displaystyle\frac{1}{m}$.

(ii)\ Given $t$ with depth $m,\ \left\{t^{\prime}\mid \rho(t, t^{\prime})<%
\displaystyle\frac{1}{m} \right\}=\{t\}$.

{\bf Remark.} Different from rational numbers, the set of all the finite
terms is not dense, but the set of all the infinite terms is dense. This
fact is stated by the above proposition. It follows from the proposition
that for any term $t\in {\rm FT},$ there is a number $m$ such that for any
term $t^{\prime},$ either $t^{\prime}=t$ or $\rho(t^{\prime},t)>\displaystyle%
\frac{1}{m}$. Therefore, ${\rm FT} \cup {\rm IT}$ is not connective. This is
one of the differences from the extension of rational numbers to real
numbers.

\bigskip Note that given two Cauchy term sequences $t^{\prime}=\{t^{\prime}_k
\mid k \in \omega\}$ and $t=\{t_k \mid k \in \omega\},$ then $%
\{\rho(t_k,t^{\prime}_k) \mid k \in \omega\}$ is a Cauchy sequence of
rational numbers, so $\displaystyle\lim_{k\rightarrow \infty}\rho(t_k,
t^{\prime}_k)$ exists.

{\bf Definition 2.4.} Given two Cauchy term sequences $t^{\prime}=\{t^{%
\prime}_k \mid k \in \omega\}$ and $t=\{t_k \mid k \in \omega\},$ we define
the distance $\rho(t,t^{\prime})=\displaystyle\lim_{k\rightarrow \infty}
\rho(t_k, t^{\prime}_k).\ t$ and $t^{\prime}$ are {\it equivalent}, denoted
by $t^{\prime}\equiv t,$ if $\rho(t^{\prime},t)=0.$

{\bf Proposition 2.5.} $\equiv$ is an equivalence relation on ${\rm FT}\cup 
{\rm IT}.$

\bigskip

{\bf Proposition 2.6.} For any set $A\subseteq {\rm FT}$ and infinitary term 
$t\in {\rm IT},$ if for any number $m$ there is a term $t^{\prime}\in A$
with $\rho(t,t^{\prime})\leq \displaystyle\frac{1}{m}$ then there is a
Cauchy term sequence $t^{\prime\prime}\subseteq A$ such that $%
t^{\prime\prime}\equiv t.$

\bigskip

We define the substitutions as in the first order logic.

{\bf Proposition 2.7.} Given a finite term $t(x),$ let $\Theta_1=\{x/r_1\},
\Theta_2=\{x/r_2\}$ be any two substitutions. Let $m$ be the least such that 
$x$ occurs in the depth $m$ of $t,$ then 
\[
\rho(t(x)\Theta_1, t(x)\Theta_2)= \displaystyle\frac{1}{m+\displaystyle\frac{%
1}{\rho(r_1,r_2)}}. 
\]
Therefore, given any finitary term $t,\ \{t\Theta_k\}$ is a Cauchy sequence if
and only if $\{\Theta_k\}$ is a Cauchy sequence, where $\{\Theta_k\}$ is a
Cauchy sequence if $\{y_k\}$ is a Cauchy sequence and $\Theta_k=\{x/y_k\}$
for every $k.$

\bigskip

Notice that for some infinitary term $t(x),\ t(r)\equiv t(r^{\prime})$ for
any terms $r$ and $r^{\prime}.$ For example, given an infinitary term $%
t(x)=\{t_k(x) \mid k \in \omega\}$ with one variable such that $m_k$ tends
to infinity, as $k$ tends to infinity, where $m_k$ is the least depth at
which $x$ occurs in $t_k,$ e.g., $t_k(x)=f^k(x)$ for some function symbol $f$%
, then $\rho(t(r),t(r^{\prime}))=0$ for any terms $r$ and $r^{\prime}.$ In
another words, when $t$ is some infinitary term satisfying certain
conditions then substituting any terms results in equivalent terms.

\bigskip

\noindent{\bf 3. A logic system $L_\IT$ for the infinitary terms}

\smallskip

Given a language ${\cal L},$ let ${\rm FT}$ and ${\rm IT}$ be the sets
defined as in the last section. 
We use $L$ to denote the first order logic on ${\cal L}.$ We assume two
basic axioms on syntax:

(3.1) The classes of variables, function symbols, predicate symbols,
constant symbols are all disjoint.

(3.2) The different variables are not equal.

\bigskip

An {\it atomic formula}\ of $L_\IT$ is in one of the following forms:

(3.3)\ $t_1\equiv t_2,$ where $t_1,t_2\in {\rm TERM}.$

(3.4)\ $p(t_1,...,t_n),$ where $p$ is an $n$-ary predicate symbol and $%
t_1,...,t_n\in {\rm TERM}.$

The formulas of $L_\IT$ are defined as in the first order logic. We assume
that the axioms and the rules of inference of $L_\IT$ are those in the first
order logic.

\bigskip

The interpretations of $L_\IT$ is defined on a structure ${\cal N}$ which is
constructed from a pre-structure ${\cal M},$ and ${\cal M}$ is a universe of
an interpretation of $L.$

A pre-structure ${\cal M}$ for $L_\IT$ is a pair ${\cal M}=<\ M,h\>$
such that $M$ is a nonempty set; $h$ is a function with ${\rm dom}%
(h)\subseteq {\cal L}; \ h(c) \in M;\ h(f): M^n\rightarrow M$ if $f$ is an $%
n $-ary function symbol; and $h(p)\subseteq M^n$ if $p$ is an $n$-ary
predicate symbol.

An assignment in ${\cal M}$ is a function $s$ such that if $x$ is a variable
symbol then $s(x) \in M.$ Given an assignment $s$ and a finitary term $t,$
we can define $s(t)$ inductively: if $t=c$ then $s(t)=h(c);$ if $t=x$ then $%
s(t)=s(x);$ and if $t=f(t_1,...,t_n)$ then $s(t)=h(f)(s(t_1),...,s(t_n)).$

Just as extending the rational numbers to the real numbers, we extend a
pre-structure ${\cal M}$ to a structure ${\cal N}=\langle N,h^{\prime}\>,$
where 
\[
N=\{\{a_k\}: \exists {t}\in {\rm IT}\exists s\forall k(a_k=s(t_k))\}/\simeq, 
\]
where $s$ is an assignment and $\simeq$ is an equivalence relation on $N$
defined as follows: Let $a=\{a_k\},$ and $a_k=s(t_k)$ for every $k,$ and $%
t=\{t_k\},$ we denote $a=s(t).$ Given any two $a=\{a_k\}$ and $b=\{b_k\},\
a\simeq b$ if there are $t, t^{\prime}\in {\rm IT}$ and an assignment $s$
such that $t\equiv t^{\prime},\ a=s(t)$ and $b=s(t^{\prime}).$ We call $%
{\cal N}$ an {\it algebraically closed extension}\ of ${\cal M}.$

For any $n$-ary predicate symbol $p,\ h^{\prime}(p)\subseteq N^n$ such that $%
h^{\prime}(p)\cap M^n=h(p).$ For any $n$-ary function symbol $f,\
h^{\prime}(f)\lceil M^n=h(f),$ and for any $a_1,...,a_n\in N,\
h^{\prime}(f)(a_1,$ $...,a_n)=\{h(f)(a_{1k},...,a_{nk}): k\in \omega\}.$

We interpret the terms and formulas in ${\cal N}$ as in the classical first
order logic. Given an assignment $s$ and a term $t,$ we can define $s(t)$
inductively: if $t$ is finitary then $s(t)$ is defined as above; if $%
t=f(t_1,...,t_n)$ and $t_1=\{t_{1k}\},...,t_n=\{t_{nk}\} \in {\rm TERM}$
then $s(t)=\{s(f(t_{1k},...,t_{nk})): k\in \omega\}.$

The truth value of a formula $\phi$ under an assignment $s$ is defined
inductively. $\phi$ is true in ${\cal N}$ under an assignment $s,$ denoted
by $\langle{\cal N},s\>\models \phi,$ if

(3.5)\ $s({t}_1)\simeq s({t}_2)$ if $\phi=t_1\equiv t_2,$

(3.6)\ $(s(t_1),...,s(t_n))\in h(p)$ if $\phi=p(t_1,...,t_n),$

(3.7)\ $\langle{\cal N},s\> \not\models \psi$ if $\phi= \lnot \psi,$

(3.8)\ $\langle{\cal N},s\>\models \psi$ and $\langle{\cal N},s\>\models
\theta$ if $\phi= \psi\land \theta,$

(3.9)\ there is an $a\in N$ such that $\langle{\cal N},s\>\models \psi(a)$
if $\phi= \exists x\psi(x).$

\bigskip

A sentence $\phi$ is satisfied in ${\cal N},$ denoted by ${\cal N}\models
\phi,$ if $\langle{\cal N},s\>\models \phi$ for any assignment $s.$ Given a
set $\Gamma$ of sentences in $L_\IT,\ {\cal N}$ is a model of $\Gamma$ if
for every $\phi\in \Gamma,\ {\cal N}\models \phi.$

\bigskip

{\bf Proposition 3.1.} (1) $L_\IT$ has stronger expressive power than the first
order logic. 

(2) $L_\IT$ is complete and compact.

{\sl Proof.} (1) Let ${\cal L}$ be the language consisting of one
function symbol $f$ and one constant symbol $c.$ The fixed point of a
function $f,$ which cannot be expressed in the first order logic, can be
expressed in $L_\IT$ by $t=\{f^k(a)\mid k\in \omega\},$ and $f(t)=t,$ where $%
f^1(a)=f(a), f^{k+1}(a)=f(f^{k}(a))$. 

(2) Just by the completeness and
compactness of first order logic, we have that $L_\IT$ is complete and compact. $\Box$
\bigskip

{\bf Remarks:} (1)\ The fixed point is expressed actually by infinitly many
terms. The main point is that we can state
some properties about the infinitary terms, just as extending the rational numbers
to the real numbers in analysis. 
 
(2) There is some research on the infinitary terms, e.g., Jaffar(1984) and Tulipani
(1994). Those studies focus on the algebraically structural or discrete properties of the infinitary 
terms. Instead, we take the whole finite or infinite terms as a continuum and focus on their 
continuous and analytic properties. 

\bigskip

\noindent{\bf 4. The continuity of predicates and functions}

\smallskip

The continuous function is a very important notion in mathematical
analysis. By the definition of the assignments, the predicate symbols and
function symbols under the interpretation have some continuous properties.
We shall give a formal definition of the continuity of the predicate and
function symbols in syntax and semantics.

\bigskip

Given an $n$-ary function symbol $f,\ f$ is {\it syntactically-continuous}\
at ${t}=({t_1},...,{t_n})$ if $\displaystyle\lim_{k\rightarrow \infty} r_k=t$
implies $\displaystyle\lim_{k\rightarrow \infty} f(r_k)=f(t).\ f$ is {\it %
syntactically-continuous}\ if $f$ is syntactically-continuous at every $t.$
By the definition of the distance, we have the following

{\bf Proposition 4.1.} Every function symbol $f$ is syntactically-continuous.

Given an $n$-ary predicate symbol $p,\ p$ is {\it syntactically-continuous}\
at ${t}=({t_1},...,{t_n})$ if there is a number $m$ such that for any
finitary term tuple $r=(r_1,...,r_n),\ \rho(r,t)<\displaystyle\frac{1}{m}$
implies $p(r)=p({t}).$

\bigskip

We now define the continuity in semantics. Given a pre-structure ${\cal M},$
let ${\cal N}=\langle N,h\>$ be its algebraically closed extension. Let $p$
be an $n$-ary predicate symbol. $h(p)$ is {\it continuous at $({t_1},...,{t_n%
})$ in ${\cal N}$ under assignment}\ $s$ if 
\[
s(p(t_1,...,t_n))=\displaystyle\lim_{k\rightarrow \infty}
s(p(t_{1k},...,t_{nk})). 
\]
$h(p)$ is {\it continuous in ${\cal N}$ under}\ $s$ if $h(p)$ is continuous
at every $({t_1},...,{t_n})$ in ${\cal N}$ under assignment $s.$

Let $f$ be an $n$-ary function symbol. $h(f)$ is {\it continuous at $({t_1}%
,...,{t_n})$ in ${\cal N}$ under assignment}\ $s$ if 
\[
s(f(t_1,...,t_n))=\displaystyle\lim_{k\rightarrow \infty}
s(f(t_{1k},...,t_{nk})). 
\]
$h(f)$ is {\it continuous in ${\cal N}$ under}\ $s$ if $h(f)$ is continuous
at every $({t_1},...,{t_n})$ in ${\cal N}$ under assignment $s.$

By the definition of the assignment, every $h(f)$ is continuous.

\bigskip

We now focus on the continuity of the predicate symbols and give a logic
system $L_c$ which is an extension of $L_\IT.$ In $L_\IT,$ for any predicate 
$p$ and infinitary term $t,$ if $p(t)$ is true under some interpretation
then there is a rational number $\delta>0$ such that for every finitary term 
$s$ with $\rho (s,t)<\delta,\ p(s)$ is true under the interpretation. We
show that such an axiomatized logic $L_c$ is sound and complete.

$L_c$ has the following axioms

(4.1)\ If $p(t)$ for $t\in {\rm TERM}$ then there is a rational number $%
\delta>0$ of form $\displaystyle\frac{1}{m}$ for some $m$ such that for any $%
r\in {\rm FT},\ \rho(r,t)\leq \delta$ implies $p(r).$

\bigskip

Given a set $\Gamma$ of formulas (theory), a proof of $\Gamma$ is a sequence 
$\{\phi_1,...,\phi_m\}$ of formulas such that for every $1\leq i\leq m,$
either $\phi_i$ is an axiom, or $\phi_i\in \Gamma$ or deducted from two
precedent formulas in the list by the inference rules. A sentence $\phi$ is
a theorem of $\Gamma,$ denoted by $\Gamma\vdash \phi,$ if there is a proof $%
\{\phi_1,...,\phi_m\}$ of $\Gamma$ such that $\phi=\phi_m.$

A sentence $\phi$ is valid in a structure ${\cal N},$ denoted by ${\cal N}%
\models \phi,$ if $\langle{\cal N},s\>\models \phi$ for any assignment $s.$
We say that $\phi$ is a logical consequence of $\Gamma$ if for every
structure ${\cal N},\ {\cal N}\models \Gamma$ implies ${\cal N}\models \phi.$

\bigskip

{\bf Theorem 4.2}(The Soundness Theorem). Given any formula $\phi$ and a
theory $\Gamma,$ if $\Gamma\vdash \phi$ then $\Gamma\models \phi.$

It can be verified routinely.

\bigskip

{\bf Theorem 4.3}(The Completeness Theorem). Given any sentence $\phi,$ if $%
\phi \models \psi$ then $\phi\vdash \psi.$ Combining the soundness theorem,
we have the following 
\[
\Gamma\vdash \phi \Leftrightarrow \Gamma\models \phi. 
\]

\bigskip

{\sl Proof.} We firstly prove the following model existence theorem. A
theory is consistent if $\Gamma\not\vdash \perp.$

{\bf Claim 4.4.} If $\Gamma$ is consistent then $\Gamma$ has a model.

{\sl The proof of the Claim.} Let $\Gamma_0=\{\sigma: \Gamma\vdash \sigma\}$
be a theory given by $\Gamma.$ Any model of $\Gamma_0$ is a model of $%
\Gamma. $

Let $\Gamma^{\prime}$ be a conservative extension of $\Gamma_0$ defined as
in the first order logic, with language ${\cal L}^{\prime}$ such that for
every sentence $\sigma\equiv \exists x\phi(x)$ there is a constant $%
c_\sigma\in {\cal L}^{\prime}-{\cal L}$ and an axiom $\exists
x\phi(x)\rightarrow \phi(c_\sigma)$ such that distinct $\sigma$'s yield
distinct $c_\sigma$'s. Then $\Gamma^{\prime}$ is a conservative extension of 
$\Gamma_0.$

Considering axiom 4.1, we prove that for any predicate $p$ and infinitary
term $t,$ if $\Gamma^{\prime}\vdash p(t)$ then there is a rational number $%
\delta$ of form $\displaystyle\frac{1}{m}$ such that for any finitary term $%
r $ with $\rho(r,t)<\delta,\ \Gamma^{\prime}\cup \{p(r)\}$ is consistent.
For the contradiction, assume that $\Gamma^{\prime}\vdash p(t)$ and for any $%
m$ there is a term $r_m$ such that $\rho(t,r_m)<\displaystyle\frac{1}{m}$
and $\Gamma^{\prime}\vdash \lnot p(r_m).$ There are two cases:

Case 4.1. $t$ does not contain any $c\in {\cal L}^{\prime}-{\cal L}.$ Since $%
\Gamma^{\prime}\vdash p(t),$ there is a finite set $\Phi\cup \Psi$ of
sentences such that $\Phi \subseteq \Gamma$ and every sentence $\sigma\in
\Psi$ has the form $\exists x\phi(x) \rightarrow \phi(c)$ for some $c\in 
{\cal L}^{\prime}-{\cal L}.$ Without loss of generality, assume that $%
\Psi=\{\sigma\}.$ Then 
\[
\begin{array}{l}
\Phi, \sigma\vdash p(t); \\ 
\Phi\vdash \sigma\rightarrow p(t); \\ 
\Phi\vdash (\exists x\phi(x)\rightarrow \phi(y))\rightarrow p(t); \\ 
\Phi\vdash \forall y[(\exists \phi(x)\rightarrow \phi(y))\rightarrow p(t)];
\\ 
\Phi\vdash (\exists x\phi(x)\rightarrow \exists y\phi(y))\rightarrow p(t);
\\ 
\Phi\vdash p(t),%
\end{array}%
\]
where $y$ is a variable that does not occur in $p(t)$ and $\Phi$. Similarly,
for any $m,$ there is a set $\Phi_m\cup \Psi_m$ of sentences such that $%
\Phi_m\subseteq \Gamma,\ \Psi_m\cap \Gamma=\emptyset$ and $\Phi_m\cup
\Psi_m\vdash \lnot p(r_m).$ If $r_m$ does not contain $c\in {\cal L}%
^{\prime}-{\cal L}$ then by the same discussion as above, $\Gamma\vdash
\lnot p(r_m).$ If $r_m$ does contain some $c\in {\cal L}^{\prime}-{\cal L}$
then 
\[
\begin{array}{c}
\Phi, \sigma_m\vdash \lnot p(r_m); \\ 
\Phi\vdash \sigma_m\rightarrow \lnot p(r_m); \\ 
\mbox{either}\ \Phi\vdash \exists x \phi(x)\rightarrow \lnot p(r_m)\ %
\mbox{or} \ \Phi\vdash \lnot \phi(c)\rightarrow \lnot p(r_m).%
\end{array}%
\]
The latter case is reduced to the former case, since $\phi(c)\in
\Gamma^{\prime}$ only if $\exists x\phi(x)\in \Gamma.$ Hence, $\Gamma \vdash
\lnot p(r_m).$ Since $c$ does not occur in $\Gamma,$ hence, there is an $%
r^{\prime}_m\in {\rm TERM}_\calL$ such that $\Gamma\vdash \lnot
p(r^{\prime}_m)$ and $r_m\equiv r^{\prime}_m.$

Case 4.2. $t$ does contain a constant symbol $c\in {\cal L}^{\prime}-{\cal L}%
.$ The discussion is the same as the second part of case 4.1.

Combining the above discussion, we have that $\Gamma\vdash p(t)$ and there
is a sequence $\{r_m\}$ or $\{r^{\prime}_m\}$ of finitary terms in ${\cal L}$
such that for every $m,\ \rho(r_m,t)<\displaystyle\frac{1}{m}$ and $%
\Gamma\vdash \lnot p(r_m).$ So $\Gamma$ is inconsistent, a contradiction.

At this moment, we extend $\Gamma^{\prime}$ to $\Gamma^{\prime\prime}$ such
that for every predicate $p$ and infinitary term $t\in {\rm TERM}_{{\cal L}%
^{\prime}},$ if $\Gamma^{\prime}\vdash p(t)$ then find the least $\delta,$
and enumerate $p(r)$ in $\Gamma^{\prime\prime}$ for any $r\in {\rm TERM}_{%
{\cal L}^{\prime}}$ with $\rho(r,t)<\delta.$ By the above discussion, $%
\Gamma^{\prime\prime}$ is consistent. Then construct the maximal consistent
extension of $\Gamma^{\prime\prime}$ into $\Gamma^{\prime\prime\prime}.$

By the maximum of $\Gamma^{\prime\prime\prime},$ we know that for any $n$%
-ary predicate symbol $p$ and terms $t_1,...,t_n,$ either $p(t_1,...,t_n)\in
\Gamma^{\prime\prime\prime}$ or $\lnot p(t_1,...,t_n)\in
\Gamma^{\prime\prime\prime}.$

We construct a pre-model ${\cal M}$ as follows: let $M=\{t\in {\cal L}%
^{\prime}: t$ is closed$\},$ and for every function symbol $f$ we define a
function 
\[
h(f)(t_1,...,t_n)=f(t_1,...,t_k); 
\]
and for every predicate symbol $p,$ we define a relation 
\[
(t_1,...,t_n)\in h(p) \Rightarrow \Gamma^{\prime\prime\prime}\vdash
p(t_1,...,t_n); 
\]
and for every constant symbol $c$ we define $h(c)=c.$

We define an equivalence relation $\sim$ on $M$ such that for any $t,s\in M,$
\[
t\sim s \Leftrightarrow \Gamma^{\prime\prime\prime}\vdash t\equiv s. 
\]
By the maximum of $\Gamma^{\prime\prime\prime},$ we can prove that $\sim$ is
an equivalence relation on ${\cal M},$ and satisfies the following claim: $%
t_i\sim s_i$ for every $1\leq i\leq n$ imply $h(p)(t_1,...,t_n)%
\Leftrightarrow h(p)(s_1,...,s_n),$ and $h(f)(t_1,...,t_n)\sim
h(f)(s_1,...,s_n).$ Hence, $\sim$ is a congruence with respect to the basic
relations and functions.

We define another structure ${\cal N}={\cal M}/\sim.$ We shall use $[t]$ to
denote the equivalence class containing $t.$ By the induction on $t$ we can
prove that i)\ if $t\equiv c$ then $[t]=[c];$ and ii)\ if $t=f(t_1,...,t_n)$
then $[t]=h(f)([t_1],...,[t_n]).$

It is routine to show that for any sentence $\phi\in
\Gamma^{\prime\prime\prime}, {\cal N}\models \phi$ if and only if $%
\Gamma^{\prime\prime\prime}\vdash \phi.$  $\Box$

\bigskip

{\bf Remarks 4.5.} (i) Given any infinitary $t,$ if there is a $\delta>0$
such that for every finitary $s, d(s,t)<\delta$ implies $p(s),$ it is
possible that $\lnot p(t).$ (ii) Given any infinitary term $t,$ if $p(t)$
then it is possible that for any $\delta>0,$ there is an term $s$ such that $%
\rho(s,t)<\delta$ and $\lnot p(s).$ That is, we do not require that for
every infinitary term $t,$ if $p(t)$ then there is a $\delta>0$ such that
for any (finitary or infinitary) term $s,\ \rho(s,t)<\delta$ implies $p(s).$

\bigskip

\noindent {\bf 5. A computation model on the infinitary terms}

Corresponding to the logic ${\cal L}_{\IT},$ we give a formal computation
model on the infinitary terms.

We shall use $\{e\}$ for natural number $e$ to denote the $e$-th Turing
machine, and $\{e\}^A$ to denote the $e$-th Turing machine with oracle $A.$
Here, $\{e\}$ can be taken as a function defined on the finitary terms of a
language ${\cal L}.$ Not every Turing computable function on ${\rm FT}$
induces a function on ${\rm TERM}.$ We shall use $\phi$ to denote a Turing
computable function from ${\rm FT}$ to ${\rm FT}$ such that $\phi$ is
continuous, i.e., for any Cauchy term sequences $t$ and $s,\ t\equiv s$
implies that $\{\phi(t_n): n\in \omega\} \equiv \{\phi(s_n): n\in \omega\}.$

{\bf Definition 5.1.} A function $f:{\rm TERM}\rightarrow {\rm TERM}$ is 
{\it computable}\ if there is a Turing computable and continuous function $%
\phi :{\rm FT}\rightarrow {\rm FT}$ such that for every $t\in {\rm TERM},$ 
\[
f(t)=\left\{ 
\begin{array}{ll}
\{\phi (t_{n}):n\in \omega \} & 
\mbox{if $\{\phi(t_n): n\in \omega\}$ is a
Cauchy term sequence} \\ 
\mbox{undefined} & \mbox{otherwise,}%
\end{array}%
\right. 
\]%
where $t\equiv \{t_{n}:n\in \omega \}.$

Such defined computable functions on ${\rm TERM}$ are not equivalent to the
partial recursive operators from $2^\omega$ to $2^\omega.$ We know that an
operator $E: 2^\omega\rightarrow 2^\omega$ is partially recursive if there
is a Turing machine $\{e\}$ such that for any $\alpha \in 2^\omega,$ 
\[
E(\alpha)(n)=\{e\}^\alpha(n) 
\]
for every $n\in \omega.$

A computable function $f$ on ${\rm TERM}$ is equivalent to the partially
computable operator $E,$ since for any $\alpha\in 2^\omega,$ 
\[
f(\alpha)(n)=\{e\}(\alpha\lceil n), 
\]
where $\alpha\lceil n$ is a restriction of $\alpha$ up to $n,$ and we take
every infinitary term as a function from $\omega$ to ${\rm FT}.$

Hence, the computable functions on ${\rm TERM}$ are something between
partially computable functions (Turing computable functions) and partially
computable operators. Namely,

{\bf Theorem 5.2.} There is a computable function on ${\rm TERM}$ which is
not a partially computable function. And every computable function on ${\rm %
TERM}$ is a partially computable operator.

\bigskip

Generally, we can define computable functions on any continuous (and
algebraically closed) ring $R$ in a similar way.

A ring $R$ is continuous if there is a subset $Q$ of $R$ such that $R$ is
the algebraical closure of $Q$ and $Q$ is countable and metric. By the
definition of the Cauchy sequences on $Q,$ we assume that the definition of
the Cauchy sequences on $Q$ can be extended to the Cauchy sequences on $R$
such that every Cauchy sequence of $R$ is equal to some element $x$ in $R,$
and every Cauchy sequence on $R$ has an equivalent Cauchy sequence on $Q.$
We can define Turing computable functions on $Q$ as usual. Not every Turing
computable functions on $Q$ can induce a function on $R.$ We shall use $%
\phi: Q\rightarrow Q$ to denote a continuous function, i.e., for any two
Cauchy sequences $x$ and $y$ on $Q,\ x\equiv y$ implies $\{\phi(x_n): n\in
\omega\} \equiv \{\phi(y_n): n\in \omega\}$ if $x=\{x_n\in Q: n\in \omega\}$
and $y=\{y_n\in Q: n\in \omega\}.$

{\bf Definition 5.3.} A function $f:R\rightarrow R$ is {\it computable}\ if
there is a Turing computable and continuous function $\phi $ on $Q$ such
that for any $x\in R,$ 
\[
f(x)=\left\{ 
\begin{array}{ll}
\{\phi (x_{n}):n\in \omega \} & \mbox{if}\ \{\phi (x_{n}):n\in \omega \}\ 
\mbox{is
a Cauchy term sequence} \\ 
\mbox{undefined} & \mbox{otherwise,}%
\end{array}%
\right. 
\]%
where $x\equiv \{x_{n}:n\in \omega \}.$

For example, Let $Q$ be the set of the rational numbers. Then $R$ is the set
of the real numbers. For any $x\in R,$ function 
\[
f(x)={\rm e}^x=1+x+\displaystyle\frac{x^2}{2!}+\displaystyle\frac{x^3}{3!}%
+\cdots +\displaystyle\frac{x^n}{n!}+\cdots 
\]
is a computable function on the real numbers. Because for any $n\in \omega,
x^{\prime}\in Q,$ 
\[
\phi(n,x^{\prime})=\left\{ 
\begin{array}{ll}
1 & \mbox{if}\ n=0 \\ 
\phi(n-1,x^{\prime})+\displaystyle\frac{x^{{\prime}n}}{n!} & \mbox{otherwise}%
\end{array}%
\right. 
\]
is computable, since $\phi(n,x^{\prime})$ can be defined recursively by the
recursive function \break $g(n,y^{\prime},x^{\prime})=y^{\prime}+\displaystyle\frac{%
x^{{\prime}n}}{n!},$ where $y^{\prime}\in Q.$ Hence, for any real number $x,\
f(x)=y,$ where $y=\{y_n: n\in \omega\},$ and $y_n=\phi(n, x_n).$

\bigskip

Both definitions of the computable functions on any continuous ring $R$ are
different from the BSS definition. In fact, both the definitions make taking
limits as the computable processes, since $\phi(x)=\displaystyle%
\lim_{n\rightarrow \infty} f(n,x\lceil n).$ Namely,

{\bf Theorem 5.4.} There is a computable function $f$ on ${\rm TERM}$ or any
continuous ring $R$ which is not BSS-computable.

\bigskip

Chadzelek and Hotz(1999) defined the analytic machines which are defined on
the real numbers and the rational numbers, and in which the infinite
computations are allowed. The computable functions defined here is different
from the analytic machines in the following point: a computable function
defined by an analytic machine produces an infinite sequence of outputs
after input; and the computable functions defined here produce infinite
sequences of outputs only after infinite sequence of inputs, and for every
input, the computation terminates.

\bigskip

\noindent{\bf 6. The real Herbrand model of Horn logic programs}

\smallskip

In this section, we will extend the Herbrand base (universe) to the real Herbrand 
base (universe) in which infinitary terms can occur, and extend 
the interpretations of the logic programs by introducing real Herbrand models of logic programs. 
Then, we will give a sufficient condition for computing 
a real Herbrand model of Horn logic programs using convergent infinite computation. 

\bigskip

{\bf Definition 6.1.} A Cauchy term sequence is called a {\it real Herbrand
term,}\  and $p(t_1,...,t_n)$ for $t_1,...,t_n\in \term$ is called a {\it 
real atom,} where $p$ is an $n$-ary predicate symbol. Define
$$\begin{array}{l}
\RHU=\IT, \\
\RHB=\{p(t_1,...,t_n)\mid p\mbox{\ is\ an\ } n\mbox{-predicate\ symbol}, 
t_1,...,t_n\in \RHU\}.
\end{array}$$
$\RHU$ and $\RHB$ are called a {\it real Herbrand universe}\ and a {\it real 
Herbrand base}, respectively.

\vspace{2mm}

{\bf Definition 6.2.} (Real Herbrand interpretation and real Herbrand model) $\tilde I \subseteq RHB$ 
is called a real Herbrand interpretation, and $M$ is called 
a real Herbrand model of a first order theory $\Gamma$ if it is a real Herbrand interpretation under which all 
sentences of the theory $\Gamma$ are true. 

\vspace{2mm}

{\bf Example 6.3.} Let $a$ be a constant symbol, $f$ a unary function symbol, $E(x, y): x=y$ and 
$\Gamma=\{\exists x E(x, f(x))\}$, then $\Gamma$ has no Herbrand model. But, $\Gamma$ has a 
real Herbrand model $\{E(\tilde b, f(\tilde b))\}$ over $RHU$, where $\tilde b$ is the infinitary term 
$\{f^k(a) \mid k \in \omega\}$.

\vspace{2mm}

There are many interesting properties for real Herbrand models of first order theories. But 
here, we will focus on Horn logic programs, not general first order theories. We will show  
how to compute a real Herbrand model of some Horn logic programs using convergent infinite computation. 

\vspace{2mm}

In the following, we recall the basic definitions in logic programming. 
A {\it literal} is an atom or the negation of 
an atom. An atom is called a {\it positive} literal whereas a  
negated atom is called a {\it negative}\ literal.
A clause $q \leftarrow q_1,...,q_m$ is called a 
{\it rule};  is called a {\it fact}, if $m=0;$ and a {\it goal}, if $q$ is 
empty. We shall use $\gamma$ to denote a clause. $q$ is called the {\it rule 
head}, denoted by $head(\gamma)$ and $q_1,...,q_m$ is called the {\it rule 
body}, denoted by $body(\gamma).$
A {\it logic program,} denoted by $\Gamma,$ is a set of clauses. 
$\Gamma$ may contain infinitely many clauses.
A clause $\gamma$ is {\it Horn} if $\gamma$ has at most one positive literal. A 
logic program is {\it Horn} if every its clause is Horn.

{\bf Definition 6.4.} A {\it real Herbrand interpretation}\ of a logic 
program $\Gamma$ is any subset $I$ of the real Herbrand base $I\subseteq 
\RHB.$ Given a logic program $\Gamma,$ a real Herbrand interpretation 
$I$ is a {\it real Herbrand model of}\ $\Gamma$ if for any substitution 
$\Theta$ and a clause $\gamma, body(\gamma)\Theta \subseteq I$ implies 
$head(\gamma)\Theta \in I.$

{\bf Example 6.5.} Suppose that there are one function symbol $f,$ one
predicate symbol $p$ and one constant symbol $a$ in $L.$ We denote the infinitary  
term $\{f^k(a) \mid k \in \omega\}$ by $f^{\infty}(a)$. Consider the logic program 
$\Gamma=\{p(f(x))\leftarrow p(x)\}$. Then, $\Gamma$ has a real Herbrand model 
$\{p(f^{\infty}(a))\}$. 

\bigskip

Let $\Gamma$ be a Horn program and define a mapping $f_\Gamma: 2^{\RHB}\rightarrow 2^{\RHB}$ by
$$f_\Gamma(A)=\{head(\gamma)\Theta\mid \gamma\in \Gamma, body(\gamma)\Theta \subseteq 
A\},$$
where $\Theta$ is any substitution.

{\bf Proposition 6.6.} (van Emden and Kowalski) If $\Gamma$ is a Horn program then $f_\Gamma$ is 
monotonic and finitary. Moreover, $f^\omega_\Gamma$ is the least real 
Herbrand model of $\Gamma,$ where $f_\Gamma^\omega$ is the least fixpoint of 
$f_\Gamma: f^0_\Gamma=\emptyset, f^1_\Gamma=f_\Gamma(\emptyset),..., 
f_\Gamma^{n+1}=f_\Gamma(f_\Gamma^n),...$

\bigskip

We define a sequence of models $\{M_k\}$ {\it convergent }\ if $\{M_k\}$ is a 
Cauchy sequence, namely, for any number $m>0,$ there is a $K$ such that for 
any $j,k\ge K,\ \rho(M_k,M_j)<\dfrac{1}{m}.$
We say that $M$ is the {\it limit}\ of a Cauchy sequence $\{M_k\},$ denoted 
by $M=\dlim_{k\rt\infty } M_k,$ if $M=\{p\in \RHB\mid \exists \{p_k\}=p\forall 
k(p_k\in M_k)\}.$ 

\bigskip
 
{\bf Lemma 6.7.} Given a Cauchy sequence $\{M_k\}$ of structures, if 
$M=\dlim_{k\rt \infty} M_k$ then $\dlim_{k\rt \infty} \rho(M_k,M)=0.$

\bigskip

In 1992, Li defined the limit of first order theory sequences in a set-theoretic way(Li, 1992). For Horn 
logic programs, we can defined such limits.

{\bf Definition 6.8.}  Given a sequence $\{\Gamma_k\}$ of Horn logic programs, we say that
$$\oo{\lim}_{k\rightarrow \infty} \Gamma_k=\ddd\bigcup_{i=1}^\infty \ddd\bigcap_{j=i}^\infty \Gamma_j$$
is the lower limit of $\{\Gamma_k\},$ and
$$\aa{\lim}_{k\rightarrow \infty} \Gamma_k=\ddd\bigcap_{i=1}^\infty \ddd\bigcup_{j=i}^\infty \Gamma_j$$
is the upper limit of $\{\Gamma_k\}.$ If $\oo{\lim}_{k\rightarrow \infty} \Gamma_k =\aa{\lim}_{k\rightarrow \infty} 
\Gamma_n$ then we say that the (set-theoretic) limit of $\{\Gamma_k\}$ exists (or $\{\Gamma_k\}$ is convergent) 
and denote it by $\LIM_{k\rightarrow \infty} \Gamma_k.$ 

\bigskip

Obviously, if $\Gamma=\displaystyle\LIM_{k\rightarrow \infty} \Gamma_k$ exists, then $\Gamma$ is still a Horn logic 
program (possibly, an infinite length Horn logic program). 

\bigskip

{\bf Problem:} We concern the following problem:
Given a sequence of Horn logic programs $\Gamma_1, \Gamma_2, \cdots, \Gamma_k, \cdots$, 
$\Gamma=\displaystyle\LIM_{k\rightarrow\infty}\Gamma_k$ exists. 
Let $M_k$ be the least model of $\Gamma_k$ for $k=1, 2, \cdots$.
The problem is if $M=\displaystyle\lim_{k\rightarrow \infty} M_k$ exists and 
is a real Herbrand model of $\Gamma=\displaystyle\LIM_{k\rightarrow\infty} \Gamma_k$?

\bigskip

The above problem does not have solutions for all sequences of Horn logic programs. 
To solve it we have to put some conditions on Horn logic programs. We assume that

{\bf Assumption 6.9.} Given a Horn logic program $\Pi $, assume that every
clause $\pi :p\leftarrow p_{1},\cdots ,p_{m}$ in $\Pi $ has the following
property: for every $i$ if $t$ is a term in $p_{i}$, then $t$ is also in $p$.

By Assumption 6.9, we would like to establish some sufficient condition for 
the above problem. We first give a fact that can be proved easily.

Given such a clause $\gamma$ satisfying assumption 6.9, by Proposition 2.7, we have the following

{\bf Proposition 6.10.} Given any clause $\gamma$ and two substitutions 
$\Theta_1,\Theta_2,$ we have that
$$\rho(head(\gamma)\Theta_1, head(\gamma)\Theta_2) \leq \rho(body(\gamma)\Theta_1, 
body(\gamma)\Theta_2).$$

{\bf Corollary 6.11.} Given an sequence $\{\Theta_n\}$ of substitutions and 
a formula $\gamma,$ if $\{body(\gamma)\Theta_n\}$ is a Cauchy sequence then 
$\{head(\gamma)\Theta_n\}$ is a Cauchy sequence.

\vspace{2mm}

{\bf Remark:} Let $\phi: body(\gamma)\mapsto head(\gamma),$ the above proposition 
can be rewritten as
$$\rho(\phi(body(\gamma))\Theta_1, \phi(body(\gamma))\Theta_2) \leq 
\rho(body(\gamma)\Theta_1, body(\gamma)\Theta_2),$$
which is an analogoue to the Lipschitz condition in dynamical systems
$$\rho(f(x_1), f(x_2)) \leq m \rho(x_1, x_2),$$
where $m$ is a constant. Here for our proposition $m$ is taken to be 1.

\bigskip

{\bf Theorem 6.12.} Given a convergent sequence $\{\Gamma_k\}$ of Horn logic 
programs. Let $M_k$ be the least real Herbrand model of $\Gamma_k$ for 
every $k.$ Then $\{M_k\}$ is a Cauchy sequence. Moreover, let 
$\Gamma=\LIM_{k\rt\infty} \Gamma_k$ and  $M=\dlim_{k\rt\infty} M_k,$ then 
$M$ is a real Herbrand model of $\Gamma.$

{\sl Proof.} We prove that $M$ is a real Herbrand model of $\Gamma,$ that 
is, given any clause $\gamma\in \Gamma$ and substitution $\Theta$ such that 
$body(\gamma)\Theta\in M,$ then $head(\gamma)\Theta\in M.$

If $\gamma$ is a fact of $\Gamma$ then there is a $K$ such that for any $k\ge 
K,\ \gamma\in \Gamma_k$ and $q=\gamma\in M_k.$ Hence, there is a Cauchy sequence 
$q_k$ such that $q_k\in M_k$ and $q=\dlim_{k\rt \infty} q_k.$

If $\gamma\in \Gamma$ is not a fact then, without loss of generality, let 
$q'=body(\gamma)\Theta \in M.$ Then there is a Cauchy seqence $\{q'_k\}$ and a 
$K$ such that $q'_k\in M_k, \gamma=\gamma_k$ for any $k\ge K$ and $q'=\dlim_{k\rt 
\infty} q'_k.$ There is a Cauchy sequence $\{\Theta_k\}$ of substitutions 
such that $q'_k=body(\gamma_k)\Theta_k$ for every $k.$ Hence, 
$\{head(\gamma_k)\Theta_k\}$ is a Cauchy sequence, and $head(\gamma_k)\Theta_k\in 
M_k$ for every $k,$ and $head(\gamma)\Theta=\dlim_{k\rt \infty} 
head(\gamma_k)\Theta_k \in M.$

\bigskip

Given any $q$ in the least real Herbrand model of $\Gamma,$ we want to show 
that there is a Cauchy sequence $\{q_k\}$ such that $q_k\in M_k$ for 
sufficiently large $k$ and $q=\dlim_{k\rt \infty} q_k.$ Let $M'$ be the 
least real Herbrand model of $\Gamma.$ We prove the claim by induction on 
$n,$ the step at which $q$ enters $M'.$ The proof is similar to the 
discussion given above. $\Box$

\bigskip

{\bf Example 6.13.}\ Assume that there are one function symbol $f,$ one
predicate symbol $p$ and one constant symbol $a$ in $L.$ We denote the infinitary  
term $\{f^k(a) \mid k \in \omega\}$ by $f^{\infty}(a)$.

Let $\{\Gamma_k\}$ be as follows: 
\[
\begin{array}{rl}
\Gamma_{1} & =\{p(f(x))\leftarrow p(x),p(f(a))\}, \\ 
\Gamma _{2} & =\{p(f(x))\leftarrow p(x),p(f^{2}(a))\}, \\ 
& \cdots \\ 
\Gamma_{k} & =\{p(f(x))\leftarrow p(x),p(f^{k}(a))\}, \\
\vdots
\end{array}%
. 
\]%
where $f^{k}(a)=f(f^{k-1}(a)),f^{1}(a)=f(a).$ Then we have 
\[
\begin{array}{rl}
M_1=f_{\Gamma_{1}}^{\omega } & =\{p(f(a)),p(f^{2}(a)),\cdots ,p(f^{n}(a)),\cdots
\}, \\ 
M_2=f_{\Gamma_{2}}^{\omega } & =\{p(f^{2}(a)),p(f^{3}(a)),\cdots
,p(f^{n}(a)),\cdots \}, \\ 
& \cdots \\ 
M_k=f_{\Gamma_{k}}^{\omega } & =\{p(f^{k}(a)),p(f^{k+1}(a)),\cdots \},\\
\vdots
\end{array}%
\]%
and 
\[
M=\displaystyle\lim_{k\rightarrow \infty }M_k=\{p(f^{\infty}(a))\}. 
\]%
Now we have 
\[
\Gamma=\displaystyle\LIM_{k\rightarrow \infty }\Gamma_{k}=\{p(f(x))\leftarrow p(x)\}, 
\]%
Therefore, $\displaystyle\lim_{k\rightarrow \infty}M_{k} \models \displaystyle\LIM_{k\rightarrow \infty }\Gamma_{k}.$ 
That is, $\displaystyle\lim_{k\rightarrow \infty}M_{k}$ is a real Herbrand model of 
$\displaystyle\LIM_{k\rightarrow \infty }\Gamma_{k}.$

\bigskip

\noindent {\bf 7. Conclusions}

\smallskip

We focus on a class of infinite computations.
First order theories are extended by using Cauchy sequences, and then a
logic for convergent infinite computations is proposed, which has stronger expressive
power than the first order logic. A class of fixed points characterizing the
logical properties of the limits can be represented by means of
infinite-length terms defined by Cauchy sequences. We show that the limit of
sequence of first order theories can be defined in terms of distance,
similar to the $\epsilon -N$ style definition of limits in real analysis.
On the basis of infinitary terms, a computation model for
convergent infinite computations is proposed.
Finally, the interpretations of logic programs are extended by introducing real 
Herbrand models of logic programs and a sufficient condition for computing 
a real Herbrand model of Horn logic programs using convergent infinite computation is given.
 
\bigskip

\noindent{\bf References:}

\smallskip

\begin{description}
\item {[1]} Abiteboul, S., Vardi, M.~Y. and Vianu, V., Computing with
Infinitary Logic. Theoretical Computer Science 149(1):101-128 (1995)

\item {[2]} Barendregt, H.~P., The Lambda Calculus, its Syntax and
Semantics, North-Holland, Amsterdam, 1984.

\item {[3]} Blum, L., Shub, M. and Smale, S., On a Theory of Computation and
Complexity over the Real Numbers: NP-completeness, Recursive Function and
Universal Machines, Bull. (New Series) Amer. Math. Soc. 21(1), 1989, 1-46.

\item {[4]} Blum, L., Cucker, F., Shub, M. and Smale, S., Complexity and
Real Computation, Springer-Verlager, 1998.

\item {[5]} Chadzelek, T. and Hotz, G., Analytic Machines, Theoretical
Computer Science, 219, 1999, 151-167.

\item {[6]} Dahr, M., Deductive Databases: Theory and Applications,
International Thomson Computer Press, 1997.

\item{[7]} Jaffar, J., "Efficient Unification over Infinite Terms", New Generation 
Computing vol. 2 no. 3, 1984, pp. 207-219

\item {[8]} Li, W., An Open Logic System, Science in China (Scientia Sinica)
(series A), {\bf 10}(1992)(in Chinese), 1103-1113.

\item {[9]} Li, W., A Logical Framework for Evolution of Specifications, in:
Programming Languages and Systems, (ESOP'94), LNCS 788, Sringer-Verlag,
1994, 394-408.

\item {[10]} Lloyd, J.~W., Foundations of Logic Programming,
Springer-Verlag, Berlin, 1987.

\item {[11]} Nienhuys-Cheng, S.~H., Distance between Herbrand
Interpretations: A Measure for Approximations to a Target Concept,
Proceedings of the 7th International Workshop on Inductive Programming,
LNAI, Springer-Verlag, 1997.

\item {[12]} Nienhuys-Cheng, S.~H., Distances and limits on Herbrand
Interpretations, Proceedings of the 8th International Workshop on Inductive
Programming, LNAI, Springer-Verlag, 1998.

\item{[13]} Tulipani, S., Decidability of the existential theory of infinite 
terms with subterm relation. Information and Computation, 108(1):1-33,
January 1994.

\item {[14]} van Emden, M.~H. and Kowalski, R.~A., The semantics of
predicate logic as a programming language, J. Association for Computing
Machinary 23(4)(1976), 733-742.

\item {[15]} Vardi, M.~Y. and Wolper, P., Reasoning About Infinite
Computations. Information and Computation 115(1):1-37 (1994).
\end{description}

\end{document}